\documentclass[12pt]{article}
\usepackage{amsmath}
\usepackage{epsfig}
\usepackage{amsfonts}
\usepackage{amssymb}

\def\beq{\begin{equation}}
\def\eeq{\end{equation}}
\def\bea{\begin{eqnarray}}
\def\eea{\end{eqnarray}}
\def\beas{\begin{eqnarray*}}
\def\eeas{\end{eqnarray*}}
\def\Tr{{\rm Tr}}

\def\XXint#1#2#3{{\setbox0=\hbox{$#1{#2#3}{\int}$}
     \vcenter{\hbox{$#2#3$}}\kern-.5\wd0}}

\begin{document}


\begin{titlepage}

\begin{centering}

\vspace*{3cm}

{\Large\bf 
 On the $N_c$-insensitivity of QCD$_{2A}$}

\vspace*{1.5cm}

{\bf Uwe Trittmann}
\vspace*{0.5cm}

{\sl Department of Physics\\
Otterbein University\\
Westerville, OH 43081, USA}

\vspace*{1cm}

\today

\vspace*{2cm}

{\large Abstract}

\vspace*{1cm}

\end{centering}

The spectrum of two-dimensional adjoint QCD is surprisingly insensitive to
the number of colors $N_c$ of its gauge group.
It is argued that the cancellation of finite $N_c$
terms is rather natural and a consequence of the singularity structure of the
theory. In short, there are no finite $N_c$ {\em contraction} terms, hence
there cannot be any finite $N_c$ singular terms,
since the former are necessary to guarantee 
well-behaved principal value integrals.
We evaluate and
categorize the matrix elements of the theory's light-cone Hamiltonian
to show how terms emerging from finite $N_c$ contributions to the
anti-commutator cancel against contributions from the purely finite $N_c$ term
of the Hamiltonian.
The cancellation is not complete; finite terms survive and modify the 
spectrum, as is known from numerical work. 
Additionally we show that several parton-number changing
  finite $N$
  matrix elements vanish. In particular, there is only one
  trace-diagonal finite $N$ correction.
It seems therefore that considering matrix elements
  rather than individual contributions can provide substantial simplifications 
  when computing the spectrum of a theory with
a large symmetry group.

\vspace{0.5cm}


\end{titlepage}
\newpage


\section{Introduction}

Two-dimensional QCD coupled to adjoint fermions (QCD$_{2A}$) continues to be of interest
since its inception several decades ago
\cite{DalleyKlebanov,BDK,Kutasov94,GrossKlebMatSmilga,GHK,Anton}
for a number of reasons. 
In particular, the screening/confinement transition as the adjoint fermions
acquire a mass has seen renewed scrutiny lately
\cite{Jacobson,Kleb21,Seifnashri,Smilga21}. 

When the fermions are {\em massless}, the theory can be formulated with
Kac-Moody currents.
This bosonization allows one to write down the dynamics as pure current-current
interactions. Then the relations to
similar theories become apparent, culminating in the discovery of a
universality of
two-dimensional theories \cite{KutasovSchwimmer}. More modest advantages
can be realized when using currents in lieu of fermions
in numerical studies. Namely, the reduced number of degrees of freedom
means reduced numerical effort \cite{UT1}. 
Meanwhile the {\em massive} theory is confining and exhibits supersymmetry
at a specific value for the fermion mass $m=\frac{g_{Y\!M}^2N}{2\pi}$.


A straightforward way to solve the theory numerically is {\em discretized light-cone
quantization} \cite{PauliBrodsky,BDK,GHK}; some of these result have very recently been
corroborated using a lattice Hamiltonian \cite{KlebLattice} and 
by standard lattice gauge theory calculations \cite{Bergner2024}.  
Some of the more recent efforts to solve for the spectrum of two-dimensional
adjoint QCD include Refs.~\cite{Kleb21,KlebFiniteN}, in which
new analytic methods
and generalization (coupling to both fundamental and adjoint fermions)
were paired with state-of-the-art sparse-matrix numerics to hone in
on the true, single-particle content of the theory. 
It seems safe to say that today the lowest states of the theory
are well described
by a straightforward DLCQ treatment. Alternative approaches include
Refs.~\cite{Katz,KatzAnand},
where conformal truncation motivates a basis-function
approach. Recently, we suggested a similarly continuous method \cite{UT5},
where instead of a single-momentum basis
a set of states completely symmetrized with respect to the theory's
automorphisms is employed,
the so-called {\em exhaustively-symmetrized light-cone
  quantization} (eLCQ).


QCD$_{2A}$ has also been studied analytically.
The authors of Ref.~\cite{Gomis}
have looked at the infrared behavior of a large class of
two-dimensional quantum field theories. The focus was on distinguishing theories with a mass
gap from gapless theories. In a later paper \cite{Gomis2}, the
renormalization flow from the ultraviolet to the infrared (IR) 
was constructed.
It is argued that adjoint QCD$_2$ 
is gapped regardless of fermion mass, and that the corresponding topological QFT in the IR has finitely
many operators.
Earlier this year, the beta functions of QCD$_{2A}$ were evaluated
in Ref.~\cite{ChermanNeuzil} to unearth the long-distance physics of the theory
treated as an effective field theory, with a special focus on the symmetry
structure of the theory and its deformations. 


The discretization of QCD$_{2A}$ works eerily well. All aspects and
behaviors seem to be correctly if crudely reproduced. 
It seems that with the latest numerical studies, including fundamental matter \cite{Kleb21} 
and finite $N$ \cite{KlebFiniteN},
one should be able to figure out how single-trace states can be multi-particle
states in the massless theory, and what a good criterion is to distinguish a true single-particle state
from loosely-bound multi-particle states.
On the other hand, this might be only understandable in a {\em Kac-Moody current} formulation.
For instance, what would a continuous Kac-Moody formulation look like,
where the modes
are arbitrarily close to one another, therefore coupling to states with infinitely many
partons? It seems that this could be a limit that is
inadequately treated in all approaches to date.
On the other hand, at least the massive spectrum seems to be treated correctly,
see Ref.~\cite{Asrat} on the heavy-fermion limit. 

We are necessarily telegraphic here, because the present, short note focuses
on but one aspect of this fascinating theory. Namely, we will probe its
insensitivity to the number of colors, aiming to explain the numerical
findings of \cite{Anton,Kleb21}.
The remainder of the paper then is organized as follows. After a brief review
of the theory and the physical states at finite $N$, we will compute
the action of the (light-cone) Hamiltonian on the states to categorize the
importance of different terms as far as their order in $N$ is concerned.
The next section looks at the consequences for the
Hamiltonian matrix elements. Namely, we will show that all trace-number
preserving interactions cancel beyond leading order. This narrows down the
possibilities of finite $N$ corrections to the spectrum to either non-singular
trace-number-violating or
parton-number-violating terms --- or both. We discuss the implication of these
results, suggest follow-up work,  and conclude.


\section{Adjoint QCD$_2$ at finite $N_c$}

Two-dimensional adjoint quantum chromodynamics is 
a non-abelian Yang-Mills theory 
coupled to fermions in the adjoint representation, and
based on the Lagrangian
\beq\label{Lagrangian}
{\cal L}=Tr[-\frac{1}{2g_{Y\!M}^2}F_{\mu\nu}F^{\mu\nu}+
i\overline{\Psi}\gamma_{\mu}D^{\mu}\Psi],
\eeq
where $\Psi=2^{-1/4}({\psi \atop \chi})$, 
and $\psi$ and $\chi$ 
being $N\times N$ matrices. The field strength tensor is
$F_{\mu\nu}=\partial_{\mu}A_{\nu}-\partial_{\nu}A_{\mu}+i[A_{\mu},A_{\nu}]$,
and the covariant derivative is $D_{\mu}=\partial_{\mu}
+i[A_{\mu},\cdot]$.
Throughout the paper,
light-cone coordinates $x^\pm=(x^0\pm x^1)/\sqrt{2}$ are used,
where $x^+$ plays the role of a time. We will work in the light-cone gauge,
$A^+=0$, thereby omitting fermionic zero modes. 

The dynamics of the system is described by the
light-cone momentum operator $P^+$ and the Hamiltonian operator $P^-$.
We express the operators in terms of the dynamic fields, namely the
right-moving adjoint fermions $\psi_{ij}$ quantized by imposing anti-commutation
relations at equal light-cone times
\beq\label{PhiCR}
\left\{\psi_{ij}(x^{-}), \psi_{kl}(y^{-})\right\} = \frac{1}{2}\,
\delta(x^{-}-y^{-})\bigl ( \delta_{il} \delta_{jk}-
{1\over N}\delta_{ij} \delta_{kl}\bigr).
\eeq
The operators are then
\beas
P^+&=&=\frac{i}{2}\int dx^- \Tr\{\psi \partial_-\psi\}\\
P^-&=&=-\frac{1}{2}\int dx^- \Tr\{im^2\psi
\frac{1}{\partial_-}\psi+\frac{g_{Y\!M}^2}{2} J^+\frac{1}{\partial^2_-}J^+\},
\eeas
with the right-moving components $J^+_{ij}=\psi_{ik}\psi_{kj}$
of the SU($N$) current.
One uses the usual decomposition of the fields in 
terms of fermion operators
\beq\label{PhiExpansion}
\psi_{ij}(x^-) = {1\over 2\sqrt\pi} \int_{0}^{\infty} dk^{+}
\left(b_{ij}(k^{+}){\rm e}^{-ik^{+}x^{-}} +
b_{ji}^{\dagger}(k^{+}){\rm e}^{ik^{+}x^{-}}\right ),
\eeq
with the anti-commutator following from Eq.~(\ref{PhiCR})
\beq\label{Commy}
\{b_{ij}(k^{+}), b_{lk}^{\dagger}(p^{+})\} =
\delta(k^{+} - {p}^{+})
\left(\delta_{il} \delta_{jk}-\frac{1}{N}\delta_{ij} \delta_{kl}\right).
\eeq
The second term of the anti-commutator needs to be included even at large $N$
because indices can conspire to yield $\sum_a\delta_{aa}=N$, but 
it will spawn many new contributions at finite $N$.
The dynamics operators are then
\begin{eqnarray}\label{ModeDecomp}
P^+ &=& \int_{0}^{\infty} dk\ k\, b_{ij}^{\dagger}(k)b_{ij}(k)\ ,\\
P^{-} &=& {m^2\over 2}\, \int_{0}^{\infty}\nonumber
{dk\over k} b_{ij}^{\dagger}(k)
b_{ij}(k) +{g_{Y\!M}^2 N\over 2\pi} \int_{0}^{\infty} {dk\over k}\
C(k) b_{ij}^{\dagger}(k)b_{ij}(k) \\
&&+ {g_{Y\!M}^2\over 4\pi} \int_{0}^{\infty} dk_{1} dk_{2} dk_{3} dk_{4}
\biggl\{ B(k_i) \delta(k_{1} + k_{2} +k_{3} -k_{4})\nonumber \\
&&\qquad\qquad\times(b_{kj}^{\dagger}(k_{4})b_{kl}(k_{1})b_{li}(k_{2})
b_{ij}(k_{3})-
b_{kj}^{\dagger}(k_{1})b_{jl}^{\dagger}(k_{2})
b_{li}^{\dagger}(k_{3})b_{ki}(k_{4})) \nonumber\\
&&\qquad + A(k_i) \delta (k_{1}+k_{2}-k_{3}-k_{4})
b_{kj}^{\dagger}(k_{3})b_{ji}^{\dagger}(k_{4})b_{kl}(k_{1})b_{li}(k_{2})
 \nonumber\\
&&\qquad + \frac{1}{2} D(k_i) \delta (k_{1}+k_{2}-k_{3}-k_{4})
b_{ij}^{\dagger}(k_{3})b_{kl}^{\dagger}(k_{4})b_{il}(k_{1})b_{kj}(k_{2})
\biggl\}, \nonumber
\end{eqnarray}
with
\begin{eqnarray}\label{HamFunctions}
A(k_i)&=& {1\over (k_{4}-k_{2})^2 } -
{1\over (k_{1}+k_{2})^2}\ , \label {ATerm} \\
B(k_i)&=& {1\over (k_{2}+k_{3})^2 } - {1\over (k_{1}+k_{2})^2 }, \nonumber\\
C(k)&=& \int_{0}^{k} dp \,\,{k\over (p-k)^2},\nonumber \\ 
D(k_i)&=& \frac{1}{(k_{1}-k_{4})^2} - \frac{1}{(k_{2}-k_{4})^2}\nonumber,
\end{eqnarray}
where the trace-number violating term $D(k_i)$ has to be included only
at finite $N$,  
the parton-number violating term is proportional to $B(k_i)$, and the
{\em contractions} $C(k)$ are necessary to render the theory finite
(``mass renormalization'').
We included this brief description of the standard way to treat the
theory on the light cone for convenience; for details see \cite{BDK,GHK}.

The crucial observation for the present work is that
the finite $N$ contributions 
to the renormalized mass, i.e. the contractions $C$, sum to zero \cite{Anton}.
On the other hand, we have new, finite $N$ contributions to the anti-commutator
and the appearance of the $D$ term.
Hence, we have to recalculate the action of the Hamiltonian
on the physical states, Eq.~(\ref{MTrFockStates}).
Note that the coefficient $\frac{g_{Y\!M}^2}{2\pi}$ of the $A,B,D$
terms is nominally down by $1/N$ relative to the mass and contraction terms.
Therefore, 
unless we gain a factor of $N$ through normalization\footnote{For instance,
  by creating two fermions, a factor $N$ emerges from
  normalization $N^{-(r+2)/2}$.}
or reoccurring indices via $\sum_a\delta_{aa}=N$, these terms are subleading.

\subsection{The States at finite $N$}

\begin{figure}[ht]
\centerline{
\psfig{file=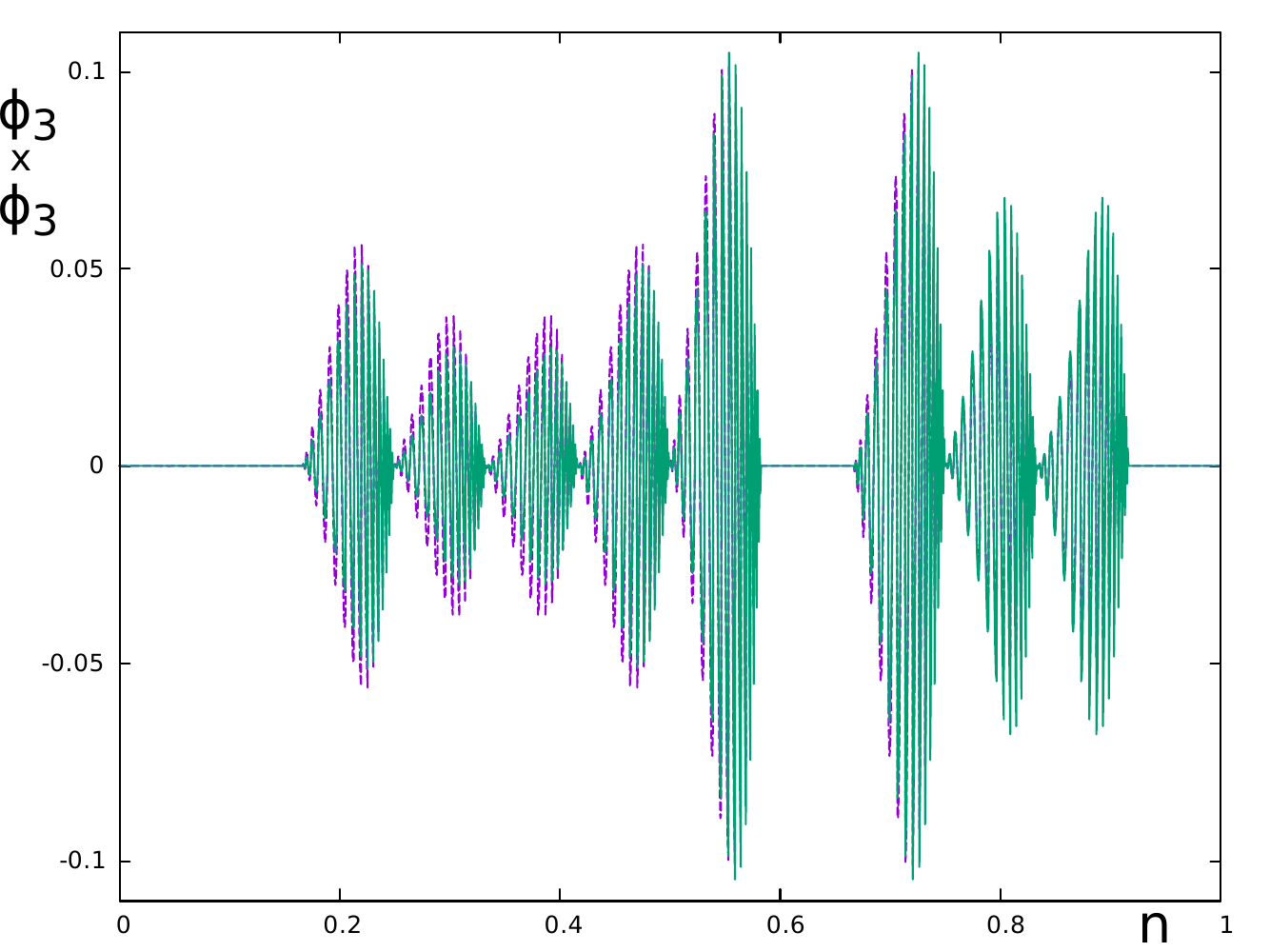,width=10cm}
}
\caption{A multi-trace state made from $\phi_3(0,0)$ and $\phi_3(2,2)$ of the massless $T_s$-odd sector as a function of
  normalized basis state number $n$. Solid, green lines: numerical, DLCQ
  calculation ($K_1=17,K_2=55$); dashed, purple lines: asymptotic eLCQ. In the intervals where
  the wavefunction vanishes, the condition, Eq.~(\ref{Kconservation}), is not
  met.}
\label{MPSPlot}
\end{figure}

The naive expression
for a state with $t$ traces with $(r_1,r_2, \cdots, r_t)=:\vec{r}$ partons
in the individual traces and $r:=\sum_{j=1}^t r_j$, namely
\beq\label{MTrFockStates}
|\Psi^{(t)}_{\vec{r}}\rangle\!=\! \frac{1}{N^{r/2}}\int_0^1 \!\!dx_1
\prod_{j=2}^{r-1} \int_0^{1-\sum_{k=1}^{j-1} x_k} \!\!\!\! dx_j
\prod_{s=1}^t\psi_{r_s}(x^{(s)}_{1},x^{(s)}_{2},\ldots, x^{(s)}_{r_{s}})
\Tr\{b^{\dagger}(x^{(s)}_{1})\cdots b^{\dagger}(x^{(s)}_{r_{s}}\}|0\rangle
\eeq
for the multi-trace states is correct, where $x^{(s)}_{r_s}$ is the $r_s$th and
therefore ``last'' momentum in the $s$th trace.
As an example,
\beas
|\Psi^{(2)}_{2,3}\rangle&=&\frac{1}{N^{5/2}} \int_0^1 dx_1\int_0^{1-x_1}
\!\!\! dx_2\int_0^{1-x_1-x_2} \!\! \!\!\!dx_3
  \int_0^{1-x_1-x_2-x_3} \!\! \!\!\! \!\!\!dx_4\,\,
\psi_{2}(x_1,x_2)\psi_{3}(x_3,x_4,x_5) \\
&&\quad\qquad\qquad\qquad\qquad\qquad\qquad
\times\Tr\{b^{\dagger}(x_{1})b^{\dagger}(x_{2})\}
\Tr\{b^{\dagger}(x_{3})b^{\dagger}(x_{4})b^{\dagger}(x_{5})\}|0\rangle
\eeas
is a fermionic two-trace, multi-particle state made from a boson and a fermion.
We can check this by comparing to
DLCQ calculations which confirm that the wavefunctions indeed factorize
into a direct product of single-trace states, see Fig.~\ref{MPSPlot}.
Note that this only works
if we use the union cHS of all unique Hilbert space cells connected
to the first one ($x_1\le x_{i}\,\,\forall i\neq 1$) by one of
the $r-1$ cyclic permutations ${\cal C}^j$ under the trace
\[
\mbox{cHS}:=\mbox{uHS}\cup{\cal C}\mbox{(uHS)}\cup{\cal C}^2\mbox{(uHS)}
+\ldots\cup{\cal C}^{r-1}\mbox{(uHS)},
\]
see \cite{UT5}. Only then
does the domain of the multi-trace sector stay the same as in the single-trace
sector. One could be worried to run into trouble since the momenta
within traces do not necessarily add up to unity.
However, trace momenta are only conserved at large $N$, and even then the only
nuisance is the sparseness of the  Hamiltonian --- since
the matrix elements between multi-trace states of
different trace momenta are zero. In DLCQ language,
\beq\label{Kconservation}
\langle K'_1, K'_2,\cdots, K'_t |\hat{H}|K_1, K_2,\cdots, K_t \rangle
=\prod_{j=1}^t \delta_{K_j}^{K'_j}. 
\eeq
The individual boundaries where
the wavefunction of a trace is required to be zero or extremal
are replaced by the factorization of the wavefunction
$\psi_r(x_1,\ldots, x_r)=\prod_{j=1}^t \psi_{r_j}(x^{(j)}_1,\ldots, x^{(j)}_{r_j})$,
i.e.~the requirement that momenta $x^{(j)}_i$ appear in one factor only.

%
%

\subsection{Notation}

While it is easy to write down the large $N$ action of the Hamiltonian on a
single-trace Fock state, $|\Psi^{(1)}_r\rangle$, Eq.~(\ref{MTrFockStates}),
and therefore the integral bound state equation, it is
cumbersome at finite $N$. There is usually no reason
to do so, because often one devises a computer algorithm to generate the
Hamiltonian matrix automatically, e.g.~in a  discretized
approach \cite{Anton,KlebFiniteN}.

For our purposes, however, it is useful to write out the Hamiltonian
matrix elements in a single-momentum Fock basis explicitly, since cancellations
can be realized {\em a priori}, i.e.~before evaluating functions arithmetically
and marveling at annulments 
in the rows and columns of a representation of the Hamilton operator. 
The disadvantage of working with matrix elements is that there are quite
a few, partly owing to the fact that the finite $N$ Hamiltonian
is not block-symmetric in the usual basis, where states have a definite
number of traces\footnote{In general, definite-trace states are
not orthogonal at finite $N$,
since relations between states exist, see the discussion
in \cite{KlebFiniteN}.}.
For instance, we can split a trace with a certain $B$ term,
Eq.~(\ref{HamFunctions}), but we
cannot join two traces with that term. Therefore, we must consider
the action on a single trace (possibly of several in a
multi-trace state) and separately the action on multiple traces ---
with different outcomes due to the different index structure.

This makes it advisable to categorize the different terms in the Hamiltonian
according to their origin ($A,B,D$), their order of $N_c$, and their parton-
and trace-(off)-diagonality.
In general, we adopt the {\em notation} ${\cal O}n$, where
${\cal O}\in\{A,B,D\}$ and $n$
is the negative of the $N_c$ exponent. For brevity, we will often
write ${\cal O}n$ for $\hat{H}_{{\cal O}n}$.
Additionally, a bar
denotes trace joining $\overline{\cal O}$, an underscore $\underline{\cal O}$ a trace-diagonal term, a tilde
a trace-splitting term $\widetilde{\cal O}$,
and a dagger a term with more creation than
annihilation operators, e.g. $\overline{B1}^{\dagger}$ is a parton-number raising
term, down $1/N$, which will decrease the number of traces.  
 
In the next section we look at the
operator structure of the terms to get a sense of which cancellations are
possible, forgetting for the moment about the
functional dependence of the interaction terms on the momenta. This line
of thought will be summarized in the block structure of the Hamiltonian in
Sec.~\ref{BlockStructure}.

\subsection{Preview of Results}

We will show that most $A$ and $D$ terms cancel completely when evaluated
in a single-momentum basis
\beas
\langle t,r|\underline{A2}+\underline{D2}|t,r\rangle =0, \qquad
\langle t,r|\widetilde{A3}+\widetilde{D3}|t,r\rangle =0,
\eeas
where $|t,r\rangle$ is a single-momentum state with $t$ traces and $r$ partons.
By construction, the are no trace-joining interactions,
$\overline{A3}=\overline{D3}=0$.
On the other hand, the next-to-leading order operators 
only cancel partially
\beas
\langle t,r|\overline{A1}+\overline{D1}|t+1,r\rangle \approx0,\qquad\qquad
\langle t+1,r|\widetilde{A1}+\widetilde{D1}|t,r\rangle \approx 0.
\eeas
Since there are {\em no contractions} at finite $N_c$, we conjecture
that the singular $1/N$ matrix elements are strictly zero between eigenstates
$|\Psi^{(t)}_{\vec{r}}\rangle$ of the full
theory
\beas
\langle \Psi^{(t)}_{\vec{r}}|\overline{A1}+\overline{D1}|\Psi^{(t+1)}_{\vec{r}}\rangle =0,\qquad\qquad
\langle \Psi^{(t+1)}_{\vec{r}}|\widetilde{A1}+\widetilde{D1}|\Psi^{(t)}_{\vec{r}}
\rangle =0.
\eeas
Our reasoning is that all $A1$ and $D1$ terms contain singularities,
which if not mutually canceled, must vanish between physical states
to render the theory finite.
Because the spectrum of the theory is known
to acquire {\em small} $1/N^2$ corrections \cite{KlebFiniteN}, those have to
come from the regular part of the $A$ terms or from the
parton-number-violating $B$ terms.
Namely, the $1/N$  $\widetilde{B1}^{\dagger}$ and
$\overline{B1}$ terms working in concert to
create a $1/N^2$ contribution proportional to
$
\langle t,r|\overline{B1}|t+1,r+2\rangle
\langle t+1,r+2|\widetilde{B1}^{\dagger}|t,r\rangle.
$


\section{Action of the Hamiltonian on the States}
\label{SecOperatorStructure}

Let's see how these calculations play out. The crucial
fact is that only 
three of the four interaction terms in Eq.~(\ref{HamFunctions}) give rise to
finite $N$ terms, namely $A(k_i),B(k_i),D(k_i)$.
The contraction term $C(k)$ does not give an extra finite $N$
term due to tracelessness \cite{Anton}.
This means that the singularities have to cancel ``internally'', as mentioned.
While the $A$ and $D$ terms
are parton-number diagonal, the $B$ terms change parton number by two.
All three of them turn out to have trace-number conserving and non-conserving
parts. Trace number can change by one with $A$ and $D$ terms,
and up to two if $B$ terms are involved.

To  realize the cancellations between interaction terms in different
sectors of the theory, we act with the Hamiltonian
on states of various trace and parton numbers.
To shorten expressions, we will often use $(x_i)$
to mean $b_{i,i+1}^{\dagger}(x_i)$, and a vacuum $|0\rangle$ at the end is understood. 

\subsubsection{Parton-number Diagonal Terms}

The parton-diagonal terms $A(k)$ and $D(k)$ have two annihilation
operators, so we get two anti-commutators with two terms each --- one leading
and the other down by $1/N$. Now, the interaction terms are already $1/N$
compared to the contraction term. So it seems we get terms proportional to
$1/N, 1/N^2,1/N^3$, unless the Kronecker deltas conspire to yield a $N$ in the
numerator. This is only possible for the $A(k)$ term, when we annihilate
two {\em adjacent} fermions in the trace. 
The $A(k)$ term's indices, Eq.~(\ref{ATerm}), then lead to the following
contributions:
\begin{itemize}
\item The large $N$ 
  contribution, where two adjacent
  fermion momenta are redistributed
  \beas
  \underline{A0}: 
 \Tr\left\{\cdots (y_i) (y_{i+1})
  \cdots\right\}
  &\longrightarrow&
  \Tr\left\{(k_3) (k_4)( y_{i+2}) \cdots (y_r)
  (y_1)\cdots (y_{i-1})\right\}.
  \eeas
\item The subleading $\frac{1}{N}$ contribution, where two nonadjacent
  fermion operators are annihilated, and then inserted adjacently in
  a second trace\footnote{Note that whenever we
  annihilate nonadjacent operators $b^{\dagger}(y_i)$ and $b^{\dagger}(y_j)$,
  we get a sign between the two sums $\sum_i(\sum_{j<i}-\sum_{j>i})$, which
  leads to the vanishing of terms symmetric under $i\leftrightarrow j$.}.
  Hence, this is a trace off-diagonal term
  \bea\label{A1}
  \widetilde{A1}&:&\,
  \Tr\left\{\cdots (y_j)\cdots (y_i)\cdots\right\}\longrightarrow
  \Tr\left\{(y_{i+1})\cdots (y_{j-1})\right\}\\
  &&\qquad\qquad\qquad\qquad\qquad\qquad
  \times
  \Tr\left\{(k_3)(k_4)(y_{j-1})(y_{j+1})(y_{j+2})
  \cdots (y_{r})\right\}\nonumber\\
  \overline{A1}&:&
   \Tr\left\{\cdots  (y_j)\cdots(y_{r_1})\right\}\Tr\left\{\cdots (\tilde{y}_i)\cdots(\tilde{y}_{r_2})\right\}
  \longrightarrow\nonumber
  \\
  &&
  \Tr\left\{(k_3)(k_4)(\tilde{y}_{i+1})\cdots(\tilde{y}_{r_2})(\tilde{y}_{1})\cdots(\tilde{y}_{i-1})(y_{j+1})\cdots(y_{r_1})(y_1)\cdots(y_{j-1})\right\}
  \nonumber
   \eea
\item {Two} different $\frac{1}{N^2}$ contributions,
  where two adjacent {\em or} nonadjacent
  fermion operators are annihilated, and then inserted adjacently in
  the same trace (trace diagonal; the indices between
  $b^{\dagger}(y_{j-1})$ and $b^{\dagger}(y_{j+1})$ are mended). There
  are two ways to fill the vacancies, which collapse into one if the
  operators are adjacent:
  \beas
  \underline{A2}&:&\,
  \Tr\left\{\cdots(y_j)\cdots (y_i)\cdots\right\} \longrightarrow
  \\
  &&
  \Tr\left\{(k_3)(k_4)(y_{i+1})\cdots (y_{j-1}) (y_{j+1})\cdots
  (y_r)(y_1)\cdots(y_{i-1})\right\}+(i\leftrightarrow j).
  \eeas
\item A $\frac{1}{N^3}$ contribution, where two adjacent
  {\em or} nonadjacent
  fermion operators are annihilated, and then ``ejected'' adjacently into
  a second trace. By construction, this is a trace splitting term only as
  indices of both vacancies are mended:
  \beas
  \widetilde{A3}:
  \Tr\left\{\cdots (y_j)\cdots (y_i)\cdots\right\}&\rightarrow&
  \Tr\left\{(k_3)(k_4)\right\}\\
&&\!\!\!\!\!\times\Tr\left\{(y_{j+1})
  \cdots (y_{i-1})(y_{i+1})\cdots (y_r)(y_1)\cdots(y_{j-1})\right\}.
  \eeas
\end{itemize}

The $D(k)$-term yields no large $N$ contribution.
Both the $1/N^2$ term $\underline{D2}$ and the $1/N^3$ $\widetilde{D3}$
term have
the same operator structure as their $A$ counterparts.
On the contrary, the $D1$ terms differ quite a bit, mapping
nonadjacent operators to nonadjacent rather than to adjacent operators.
These newly created operators end up in {\em separate} traces in $\widetilde{D1}$
and annihilating operators in different traces will join traces in $\overline{D1}$
\bea\label{D1}
\widetilde{D1}&:& \Tr\left\{\cdots(y_j)\cdots (y_i)\cdots\right\}
\longrightarrow
\Tr\left\{(k_3)(y_{j+1}) \cdots (y_{i-1})\right\}\\
&&\qquad\qquad\qquad\qquad\qquad\qquad
\times \Tr\left\{(k_4)(y_{i+1})\cdots (y_r)(y_1)\cdots(y_{j-1})\right\}\nonumber\\
\overline{D1}&:&
  \Tr\left\{\cdots  (y_j)\cdots(y_{r_1})\right\}\Tr\left\{\cdots (\tilde{y}_i)\cdots(\tilde{y}_{r_2})\right\}
  \longrightarrow\nonumber
  \\
  &&
 \quad \qquad\Tr\left\{(k_3)(y_{j+1})\cdots(y_{r_1})(y_1)\cdots(y_{j-1})(k_4)(\tilde{y}_{i+1})\cdots(\tilde{y}_{r_2})(\tilde{y}_{1})\cdots (\tilde{y}_{i-1})\right\}.
  \nonumber
\eea

\subsubsection{Parton-number Changing Terms}

This leaves the parton-number changing terms $B(k_i)$, which are somewhat
confusing, since they are parton number off-diagonal as well as potentially
trace off-diagonal. Furthermore, they allow for different sources of factors
$N_c$: from the anti-commutator's finite $N$ part, decreasing the power of
$N_c$, and also from ``adjacency'' which can {\em increase} the power of $N_c$.
Besides, the number of traces can vary by up to two, whereas the
parton-diagonal interactions can only change trace-number by one.
On the other hand, we cannot say much about these terms on physical grounds,
since they are not singular. We include them here mostly for
completeness.

The pair-production terms include the leading large $N$ term
$\underline{B0}^{\dagger}$ (insert three adjacent
partons where one is taken out) and the subleading, trace-splitting
term $\widetilde{B1}^{\dagger}$
(take one out and mend the vacancy, create a second trace with three fermion
operators).
Clearly, $\underline{B0}^{\dagger}$ is trace diagonal.
There is no term $\overline{B1}^{\dagger}$, since taking out just one operator
can't possible join traces. Note that this leads to an asymmetric Hamiltonian
in the naive basis.

The two-parton annihilation terms ${B}n$ are naturally down
$1/N$ due to the normalization factor from the annihilated
operators. Therefore
the trailing contribution with an additional $1/N^3$ from three finite $N$
contributions to the anti-commutators would be down $1/N^5$. Alas,
it vanishes due to tracelessness. The large $N$ term $\underline{B0}$
takes three adjacent operators out and replaces them
with one of same total momentum, so it is trace diagonal.
Furthermore, the adjacent Kronecker deltas yield a factor $N^2$, so overall
it has a factor $N$ --- which is part of the 't Hooft coupling $\frac{g_{Y\!M}^2N}{2\pi}$.
The term $b^{\dagger}bbb$ is
usually $1/N^2$ ($1/N$ since there is no factor $N$ with
$g$, and $1/N$ from norm) and trace off-diagonal. However, this is true only
if the annihilated
operators all sat apart, whence we split into {\em three} traces
\[
\widetilde{B2}_{3\Tr}: \Tr\{(y_1)\cdots (y_{u-1})(k_4)(y_{s+1})\cdots (y_r)\}
\Tr\{(y_{u+1})\cdots (y_{w-1})\}\Tr\{(y_{w+1})\cdots (y_{s-1})\}.
\]
Otherwise, two traces and an $N$ emerge, so this term is down $1/N$
\[
\widetilde{B1}: 
\Tr\{(y_1)\cdots (y_{u-1})(k_4)(y_{w+1})\cdots (y_{s-1})(y_{s+1})\cdots (y_r)\}
\Tr\{y_{u+1}\ldots y_{w-1}\}.
\]
There is another $1/N^2$ term, $\underline{B2}_{ac}$,
with a $1/N$
anti-commutator term, which is trace diagonal and gains a power of $N$
because of adjacency
of two annihilated operators:
{\small
\[
\underline{B2}_{ac}: \Tr\{\cdots (y_{u})(y_{u+1})\cdots (y_{s})\cdots\}
\rightarrow \Tr\{(y_1)\cdots (y_{u-1})(k_4)(y_{u+2})\cdots (y_{s-1})(y_{s+1})\cdots (y_r)\}.
\]
}
However, if applying to non-adjacent operators, we get a trace-splitting $1/N^3$
term 
\[
\widetilde{B3}: \Tr\{(y_1)\cdots (y_{u-1})(k_4)(y_{w+1})\cdots(y_{s-1})(y_{s+1})\cdots (y_r)\}
\Tr\{(y_{u+1})\cdots (y_{w-1})\}.
\]
There is yet another possibility to create a $1/N^2$ term: if no $1/N$
commutator term is present and the operators act such that adjacent
indices do not create an extra $N=\sum_a\delta_{aa}$ (operators sit on
``wrong side''). Let's call this term $\underline{B2}_{ws}$ and subsume it
by defining $\underline{B2}:=\underline{B2}_{ac}+\underline{B2}_{ws}$.
The  $1/N^4$ contribution $\underline{B4}$ with two finite $N$
anti-commutator contributions
is straightforward: annihilate three (not necessarily adjacent) operators,
mend two vacancies and fill in with the new operator. Terms have the structure 
\beas
\underline{B4}&:&
\Tr\left\{\cdots(y_u)\cdots (y_w)\cdots (y_s)\right\}\longrightarrow\\
&&\!\!\!\!
\Tr\left\{(y_u+y_w+y_s)(y_{u+1})
\cdots (y_{w-1}) (y_{w+1})\cdots
(y_{s-1}) (y_{s+1})\cdots (y_r)(y_1)\cdots(y_{u-1})\right\}
\eeas
plus permutations of $(u,w,s)$.
On to the terms which join traces.
Since $\underline{B0}$,  $\underline{B2}$, and $\underline{B4}$ are
trace diagonal, we have three terms with the following structure:
{\small
\beas
\overline{B1}
&:& \Tr\{(y_1)\ldots (y_{r_1})\}\Tr\{(\bar{y}_{1})\ldots (\bar{y}_{r_2})\}\longrightarrow \\
&& \Tr\{(y_1)\ldots (y_{u-1})(k_4) (\bar{y}_{s+1})\ldots (\bar{y}_{r_2})
(\bar{y}_{1})\ldots (\bar{y}_{s-1})({y}_{u+2})\ldots ({y}_{r_1})\},\\
\overline{B2}_{3\Tr}&:&
\Tr\{(y_1)\ldots (y_{r_1})\}\Tr\{(\bar{y}_1)\ldots (\bar{y}_{r_2})\}
\Tr\{(\tilde{y}_1)\ldots (\tilde{y}_{r_3})\}\longrightarrow 
 \Tr\{(y_1)\ldots (y_{u-1})\\
 &&\quad\cdot(k_4)(\tilde{y}_{s+1})\ldots (\tilde{y}_{r_3})(\tilde{y}_1)\ldots (\tilde{y}_{s-1})(\bar{y}_{w+1})
 \ldots (\bar{y}_{r_2})(\bar{y})_1)\ldots(\bar{y}_{w-1})(y_{u+1})\ldots (y_{r_1})\},
 \\
\overline{B3}
&:& \Tr\{(y_1)\ldots (y_{r_1})\}\Tr\{(\bar{y}_{1})\ldots (\bar{y}_{r_2})\}
\longrightarrow\\
&& \Tr\{(y_1)\ldots (y_{u-1})(k_4) (\bar{y}_{w+1})\ldots (\bar{y}_{r_2})
(\bar{y}_{1})\ldots (\bar{y}_{w-1})({y}_{u+1})\ldots({y}_{s-1})({y}_{s+1})
\ldots(y_{r_1})\}.  
\eeas
}

\subsection{The Block Structure of the Hamiltonian}
\label{BlockStructure}

From the discussion above we glean the trace-block structure of the Hamiltonian 
\beq\label{HamStructure}
(\hat{H})=\left(
\begin{array}{ccc}
  H_{\underline{{\cal O}024}} &  H_{\overline{{\cal O}13}} & H_{\overline{{\cal O}2}}\\
  H_{\widetilde{{\cal O}13}} & H_{\underline{{\cal O}024}} &  H_{\overline{{\cal O}13}} \\
  H_{\widetilde{{\cal O}2}} & H_{\widetilde{{\cal O}13}} & H_{\underline{{\cal O}024}} \\
\end{array}
\right),
\eeq
with
{\footnotesize
\[
H_{\underline{{\cal O}024}} \!=\!
\left(\!\!
\begin{array}{ccc}
  \underline{A0}\!\!+\!\!\underline{A2}\!\!+\!\!\underline{D2} &\underline{B0}\!\!+\!\!\underline{B2}\!\!+\!\!\underline{B4} & 0 \\
\underline{B0}^{\dagger} & \underline{A0}\!\!+\!\!\underline{A2}\!\!+\!\!\underline{D2} &\underline{B0}\!\!+\!\!\underline{B2}\!\!+\!\!\underline{B4}\\
0& \underline{B0}^{\dagger} & \underline{A0}\!\!+\!\!\underline{A2}\!\!+\!\!\underline{D2}
  \end{array}\!\!\right),
H_{\overline{{\cal O}13}} \!=\!
\left(\!\!
\begin{array}{ccc}
  \overline{A1}\!\!+\!\!\overline{D1}&\overline{B1}
  \!\!+\!\!\overline{B3}& 0 \\
  0 & \overline{A1}\!\!+\!\!\overline{D1} &\overline{B1}
  \!\!+\!\!\overline{B3}\\
0& 0 & \overline{A1}\!\!+\!\!\overline{D1}
  \end{array}\!\!\right),
\]
}
{\footnotesize
\[
H_{\overline{{\cal O}2}} =
\left(\!
\begin{array}{ccc}
0& \overline{B2}_{3Tr}&0\\  
0&0& \overline{B2}_{3Tr}\\
0&0&0
\end{array}\!
\right),
H_{\widetilde{{\cal O}13}} =
\left(\!
\begin{array}{ccc}
  \widetilde{A1}\!\!+\!\!\widetilde{D1}\!\!+\!\! \widetilde{A3}
  \!\!+\!\!\widetilde{D3}&\widetilde{B1}\!\!+\!\!\widetilde{B3} & 0 \\
  \widetilde{B1}^{\dagger}& \widetilde{A1}\!\!+\!\!\widetilde{D1}
  \!\!+\!\! \widetilde{A3}\!\!+\!\!\widetilde{D3}&\widetilde{B1}
  \!\!+\!\!\widetilde{B3} \\
  0& \widetilde{B1}^{\dagger}  & \widetilde{A1}\!\!+\!\!\widetilde{D1}
  \!\!+\!\! \widetilde{A3}\!\!+\!\!\widetilde{D3}
  \end{array}\!
\right)\!\!,
\]
}
and $H_{\widetilde{{\cal O}2}}$ has the same structure as $H_{\overline{{\cal O}2}}$,
only that the elements are trace splitting instead of trace joining.
For brevity, we limited the blocks $H_{\cal O}$ to three rows and columns,
which represent $r,r+2,r+4$ parton blocks. In reality, these are infinite
matrices.
Note that
the $Bn$ always appear above the diagonal of a block
due to pair annihilation. Also note that 
$\overline{A3}$ and $\overline{D3}$ vanish. Hence, the Hamiltonian matrix is
not (block-)symmetric in the naive, definite-trace-number basis,
as remarked earlier.

We will now proceed to show that $\underline{A2}+\underline{D2}=0$ and 
$\widetilde{A3}+\widetilde{D3}=0$, as well as the vanishing of the matrix
elements of $\widetilde{B3},\overline{B3}$, and $\underline{B4}$,
thereby simplifying the Hamiltonian.
This also makes more apparent that matrix elements destroying the symmetry
of the Hamiltonian in the naive basis are down at least $1/N^2$,
which helps explain why early numerical studies \cite{Anton} could ignore the
finite $N$ corrections to the norms of states, i.e.~violations of
orthogonality.

\section{Evaluating the Matrix Elements}

We now project onto the 
outgoing states $|\vec{x}^{(r')}\rangle$ to complete the evaluation
of the matrix elements. Note that the $|\vec{x}^{(r')}\rangle$ are the naive
states, i.e.~the states forming a basis at large $N$. In this note, we shall
ignore the fact that the overlap between these states in general does not
vanish  at finite $N$.
The relations between states were worked out in \cite{KlebFiniteN}, and
must be implemented in a full (eLCQ) treatment of the theory, which we
leave to future work.

After some algebra, the trace-diagonal $1/N^2$ matrix elements turn out to be
{\small
\bea\label{MatrixElementsA2D2}
\langle \vec{x}^{(r')}|\underline{A2}+\underline{D2}|\vec{y}^{\,(r)}\rangle
&=&\frac{g_{Y\!M}^2}{2\pi N}\delta^{r'}_r\sum_{i'=1}^{r'}\sum_{i=1}^r
\left(\sum_{j=1}^{i-1}-\sum_{j=i+1}^{r}\right)(-1)^{i+j}
\delta(x_{i'}+x_{i'+1}-y_i-y_j)\nonumber\\
&&\!\!\!\!\!\!\!\!\!\!\!\!\!\!\!\!\!\!\!\!\!\!\!\!\!\!\!\!\!\!\!\!
\!\!\!\!\!\!\!\!\!\!\!\!\!\!\!\!
\times\Big[\delta\left(\vec{x}_{Sp}^{\,(i',i'+1)}-\vec{y}_{Sp}^{\,(j,i)}\right)
  (-1)^{(r'-1)[i-i'-\theta(i-j)]} + (i\leftrightarrow j)
  \Big]\\
&&\!\!\!\!\!\!\!\!\!\!\!\!\!\!\!\!\!\!\!\!\!\!\!\!\!\!\!\!\!\!\!\!
\!\!\!\!\!\!\!\!\!\!\!\!\!\!\!\!
\times\left[
  \left(\frac{1}{(x_{i'}-y_i)^2}-\frac{1}{(x_{i'}+x_{i'+1})(y_{i}+y_{j})}\right)
  +\frac{1}{2}\left(\frac{1}{(x_{i'}-y_j)^2}
  -\frac{1}{(x_{i'}-y_i)^2}\right)\right]=0,  \nonumber
\eea
}
\noindent where $\vec{y}_{Sp}^{\,(n,m)}$ represents a cyclically fixed
string of spectator
momenta where the $n$th and $m$th momenta have been erased
and which starts with the $m+1$th momentum\footnote{The signs
  are due to annihilating the $i$th
  and $j$th operator 
  and the relative cyclic
  rotation yielding $(-1)^{(i'-i)(r'-1)}$ if $j>i$ and $(-1)^{(i'-i-1)(r'-1)}$
  otherwise for states with even parton number $r'$.
}, whereas $\theta(x)$ is the step function.

Hence, after carefully taking into account the signs, one arrives at the astonishing
result that {\bf the four terms in the last row exactly cancel} when the
  sums are executed!
  In essence, this is due to the sign change in $D(k_1,k_2)=-D(k_2,k_1)$,
  Eq.~(\ref{HamFunctions}),
  being compensated by the signs in Eq.~(\ref{MatrixElementsA2D2}),
resulting in the last two terms canceling the first, singular term. 
The second, regular term $(x_{i'}+x_{i'+1})^{-2}=(y_{i}+y_{j})^{-2}$
vanishes due to the relative sign between the sums over $j$.
In hindsight, this cancellation could have been predicted by the
requirement that the physics of bound states has to pan out.
Namely, as was noted as early as \cite{Anton}, the {\em contractions}
do not contribute at finite $N$ due to tracelessness. Therefore the cancellation
of singularities must come from the finite $N$ term, $D(k_i)$,
absent at large $N$. This in turn means that some of the finite $N$ terms
are out of the game: the existence of finite bound states requires
the theory to be largely insensitive to $N$! The cancellation is easily checked (and overlooked)
with a finite $N$ DLCQ code; one has to suppress the large $N$ contributions.

We now have to show that $N$ dependence does not creep back in on the off-diagonals.
However, it should be clear that this cannot be the case for the same reason:
the $A(k_i)$ terms have a singular momentum dependence, which must be canceled
by their $D(k_i)$ counterparts -- be it on the diagonal or off.

The $1/N$ matrix element from the $A$ term is
{\small
\bea\label{MatrixElementA1}
\langle \vec{x}^{(s_1)}|\langle\vec{x}^{(s_2)}|\widetilde{A1}|\vec{y}^{\,(r)}\rangle
&=&\frac{g_{Y\!M}^2}{2\pi}\sum_{t=1}^2\delta^{s_1+s_2}_r(-1)^{s_ts_{3-t}}\sum_{i'=1}^{s_t}\sum_{j'=1}^{s_{3-t}}
\sum_{i=1}^r\left(\sum_{j=1}^{i-1}-\sum_{j=i+1}^{r}\right)(-1)^{i+j}\nonumber\\
&&\!\!\!\!\!\!\!\!\!\!\!\!\!\!\!\!\!\!
\times(-1)^{s_t(i-i')+s_{3-t}(j-j')}(-1)^{(r-max(i,j))(j-\theta(i-j))}\nonumber\\
&&\!\!\!\!\!\!\!\!\!\!\!\!\!\!\!\!\!\!
\times
\delta\left(\vec{x}_{Sp}^{(s_t; i')}-\vec{y}^{\,[j+1,i-1]}_{Sp}\right)
\delta\left(\vec{x}_{Sp}^{(s_{3-t}-2; j')}-\vec{y}^{\,[i+1,j-1]}_{Sp}\right)\nonumber\\
&&\!\!\!\!\!\!\!\!\!\!\!\!\!\!\!\!\!\!
\times\delta\left(x^{(3-t)}_{j'}+x^{(3-t)}_{j'+1}-y_i-y_j\right)
\left(\frac{1}{(y_i+y_j)^2}-\frac{1}{\left(x^{(3-t)}_{j'+1}-y_j\right)^2}\right),
\eea
}
where $\vec{x}_{Sp}^{\,(s_t;i')}:=(x^{(t)}_{i'},x^{(t)}_{i'+1},\ldots x^{(t)}_{s_t},
x^{(t)}_1,\ldots,x^{(t)}_{i'-1})$  with $s_t$ momenta and
$\vec{x}_{Sp}^{\,(s_t-2;j')}:=(x^{(t)}_{j'+2},x^{(t)}_{j'+3},\ldots x^{(t)}_{s_t},
x^{(t)}_1,\ldots,x^{(t)}_{j'-1})$ with $s_t-2$ spectator momenta. In
\beq\label{yspectator}
\vec{y}_{Sp}^{\,[i+1,j-1]}:=
\left\{
(y_{i+1}\ldots,y_{r-1},y_r,y_1,y_2\ldots y_{j-1})\quad; \quad i>j
\atop
(y_{i+1},\ldots, y_{j-1}) \quad; \quad j>i
\right.
\eeq
we have $r-i+j-1$ and $j-i-1$ momenta, respectively.
Analogously we compute the $\widetilde{D1}$ matrix element
{\small
\bea\label{MatrixElementD1}
\langle \vec{x}^{(s_1)}|\langle\vec{x}^{(s_2)}|\widetilde{D1}|\vec{y}^{\,(r)}\rangle
&=&\frac{g_{Y\!M}^2}{2\pi}\sum_{t=1}^2\delta^{s_1+s_2}_r(-1)^{s_ts_{3-t}}\sum_{i'=1}^{s_t}\sum_{j'=1}^{s_{3-t}}
\sum_{i=1}^r\left(\sum_{j=1}^{i-1}-\sum_{j=i+1}^{r}\right)(-1)^{i+j}\nonumber\\
&&\!\!\!\!\!\!\!\!\!\!\!\!\!\!\!\!\!\!
\times(-1)^{s_t(i-i')+s_{3-t}(j-j')}(-1)^{[r-i+\theta(j-i)][i-\theta(i-j)]+j-1
  -r\theta(j-i)}\nonumber\\
&&\!\!\!\!\!\!\!\!\!\!\!\!\!\!\!\!\!\!
\times
\delta\left(\vec{x}_{Sp}^{[s_t;i']}-\vec{y}_{Sp}^{\,[i+1,j-1]}\right)
\delta\left(\vec{x}_{Sp}^{[s_{3-t};j']}-\vec{y}_{Sp}^{\,[j+1,i-1]}\right)
\nonumber\\
&&\!\!\!\!\!\!\!\!\!\!\!\!\!\!\!\!\!\!
\times\delta\left(x^{(t)}_{i'}+x^{(3-t)}_{j'}-y_i-y_j\right)
\left(\frac{1}{\left(x^{(t)}_{i'}-y_i\right)^2}
-\frac{1}{\left(x^{(t)}_{i'}-y_j\right)^2}\right),
\eea
}
where $\vec{x}_{Sp}^{[s_t;i']}:=(x^{(t)}_{i'+1},x^{(t)}_{i'+2},\ldots x^{(t)}_{s_t},
x^{(t)}_1,\ldots,x^{(t)}_{i'-1})$
with $s_t-1$ spectator momenta.
Here, the factor $\frac{1}{2}$ is absent because it
is compensated by the $D1$ term producing twice as many contributions.
This is by no means obvious --- one has to go carefully through all the sums.
Both the signs and the spectator delta functions differ
in the expressions for the $\widetilde{A1}$ and $\widetilde{D1}$  matrix
elements, and 
these $1/N$ terms do {\em not} cancel completely.
This is due to the different index
structure of Eqs.~(\ref{A1}) and (\ref{D1}). In particular, the regular
part of $\widetilde{A1}$ does not cancel.

The $1/N^3$ contributions of the $\underline{A3}$ and $\underline{D3}$
terms
have exactly the same operator structure, see Sec.~\ref{SecOperatorStructure}.
Therefore we get to apply the same logic as for the cancellations of
$\underline{A2}$
and $\underline{D2}$, and we conclude
\beas
\langle \vec{x}^{(2)}|\langle
\vec{x}^{(r'-2)}|\underline{A3}+\underline{D3}|\vec{y}^{\,(r)}\rangle&=&0.  
\eeas
Recall that the associated trace-joining terms vanish {\em ab ovo}.

While we have no physical reason to suspect that the matrix elements
of the parton-number
violating $B$-terms
vanish, it turns out that some do. These cancellations must
then follow from the (internal) symmetries of the theory, and are interesting
in their own right.
While the matrix elements of $B1$ and $B2$ do not vanish in general, they
become less important as the harmonic resolution grows in a DLCQ formulation.
Namely, we find that
the relative number of non-zero $B1$ matrix elements
decreases as $K$ grows: $K=13,14,15,16$ has $7\%,6\%,5\%,4\%$
$B1$ transitions. It seems possible that in the continuum limit, finite
$N$ contributions to pair
production become unimportant.

After some algebra, the trace-diagonal, parton-number changing
$1/N^4$ matrix elements turn out to be zero
{\small
\bea\label{MatrixElementB4}
\langle \vec{x}^{(r')}|\underline{B4}
|\vec{y}^{\,(r)}\rangle
&\!\!=\!\!&\frac{g_{Y\!M}^2}{2\pi N^3}\delta^{r'}_{r-2}\sum_{i'=1}^{r'}\sum_{i=1}^r
\left[\sum_{j=1}^{i-1}\left(\sum_{k=1}^{j-1}\!-\!\!\sum_{k=j+1}^{i-1}\!\!+\!\!\sum_{k=i+1}^{r}\right)
  \!-\!\!\sum_{j=i+1}^{r}\left(\sum_{k=1}^{i-1}\!-\!\!\sum_{k=i+1}^{j-1}\!\!+\!\!\sum_{k=j+2}^{r}\right)
  \right]\nonumber\\
  &&\!\!\!\!\!\!\!\!\times
  (-1)^{i+j+k-1+(i'-1)(r'-1)}
\delta(x_{i'}-y_i-y_{j}-y_k)\left(\frac{1}{(y_{j}+y_i)^2}
  -\frac{1}{(y_{k}+y_{j})^2}\right)
 \nonumber\\
&&\!\!\!\!\!\!\!\!\times
 \left[\delta\left(\vec{x}_{Sp}^{\,[r',i']}-\vec{y}_{Sp}^{\,(i,j,k)}\right)
   +\delta\left(\vec{x}_{Sp}^{\,[r',i']}-\vec{y}_{Sp}^{\,(j,k,i)}\right)
   +\delta\left(\vec{x}_{Sp}^{\,[r',i']}-\vec{y}_{Sp}^{\,(k,i,j)}\right)\right]=0, 
 \eea
 }
where $\vec{y}_{Sp}^{\,(n,m,k)}$ represents a cyclically fixed
string of spectator
momenta where the $n$th, $m$th and $k$th momenta have been erased
and which starts with the $m+1$th momentum.
The mutual cancellation of terms is rather subtle, resting both
on the symmetries of the interaction kernel\footnote{Obviously,
  $B(y_i,y_k)=-B(y_k,k_i)$, but it also fulfills the ``{\em Jacobi identity}\,''
  $B(i,j,k)+B(j,k,i)+B(k,i,j)=0$.} $B(\vec{y}_{\overline{Sp}})$,
and the signs of summands.
Note that the
individual contributions do not vanish --- only their sum does.

We can shed some more light on this by studying the vanishing of
$\widetilde{B3}$
in more detail. This trace-off-diagonal, PNV $1/N^3$ matrix
elements is
\bea\label{MatrixElementB3}
\langle \vec{x}^{(s_1)}|\langle \vec{x}^{(s_2)}|\widetilde{B3}|\vec{y}^{\,(r)}\rangle
&\!\!=\!\!&\frac{3 g_{Y\!M}^2}{\pi N^2}\delta^{s_1+s_2}_{r-2}\sum_{t=1}^{2}
\sum_{i'=1}^{s_t}\sum_{j'=1}^{s_{3-t}}\sum_{i=1}^r
\left[\sum_{j=1}^{i-1}\left(\sum_{k=1}^{j-1}+\sum_{k=i+1}^{r}\right)
  \!+\!\!\sum_{j=i+1}^{r}\sum_{k=i+1}^{j-1}
  \right]\nonumber\\
&&
\times
  (-1)^{s_ts_{3-t}+i+k-1+(i'-1)(s_t-1)+(j'-1)(s_{3-t}-1)}
\delta(x_{i'}-y_i-y_{j}-y_k)
 \nonumber\\
&&
\times
\delta\left(\vec{x}_{Sp}^{[s_t; i']}-\vec{y}^{\,[j+1,k-1;i]}_{Sp}\right)
\delta\left(\vec{x}_{Sp}^{(s_{3-t}; j')}-\vec{y}^{\,[k+1,j-1]}_{Sp}\right)\nonumber\\
&&\times\left(\frac{1}{(y_{j}+y_i)^2}
  -\frac{1}{(y_{k}+y_{j})^2}\right)=0.
\eea
where $\vec{y}_{Sp}^{\,[i,j;k]}$ is a generalization of the previously defined
$\vec{y}_{Sp}^{\,[i,j]}$, Eq.~(\ref{yspectator}),
in that the $k$th momentum has been erased from the latter.
We cyclically rotated the state $j$ times before
annihilations and associated index alterations, which explains why only
$i,j$ appear in the sign part.
es we get a cancellation of
The factor three 
comes from symmetries of expressions. Namely,
the three terms stemming from attaching the $1/N$ factor to one of the three
annihilators can be mapped into each other. 
Each term has $3!$ sums due to the ordering of three indices, and labeling
the 18 expressions $T_{mn}$, $m=1,2,3$, $n=1,\ldots, 6$, we have, e.g.
$T_{11}(w\leftrightarrow u, s_1\leftrightarrow s_2)= T_{14}$,
and $T_{21}(w\leftrightarrow s, s_1\leftrightarrow s_2)= T_{14}$,
where $s_1\leftrightarrow s_2$ means swapping the traces
which can yield a sign.
This allows for the elimination of half the sums and two thirds of the
terms in the last line relative to the similar expression for $\underline{B4}$,
Eq.~(\ref{MatrixElementB4}).
Now, the interaction kernel has its own symmetries, as already remarked. 
In total, these automorphisms ensure that the matrix element, i.e.~the
sum over all contributions, vanishes.

\subsection{The Simplified Block Structure of the Hamiltonian}
\label{SimpleBlockStructure}

We have now seen that for individual reasons most finite $N$
corrections ($\underline{A2}+\underline{D2},\widetilde{A3}+\widetilde{D3},
\underline{B3},\underline{B4}$) vanish, while others
($\widetilde{B1},\underline{B2}$)
do not --- at least at the matrix element level.
One has to suspect that there is an overarching principle,
a symmetry at work here.
Alas, we have not been able to fully figure it out.
What we can say is what prevents the large $N$ terms from canceling
is the constraint of adjacency of operators, which eliminates the
possibility of partner terms with equivalent momenta but opposite
sign. Additionally, we note that while the $B$ and $D$ kernels fulfill
the ``{\em Jacobi identity}\,'', the $A$ kernel does not, because its mixing of
momentum sum and difference.  
These rather curious observations are far reaching.  
Namely, we learn that {\em considering matrix elements
  rather than individual contributions can provide substantial simplifications 
  when computing the spectrum of a theory with
a large internal symmetry group.}
In particular, there are
no matrix elements proportional to $1/N^3$ nor to  $1/N^4$, so
$H_{\underline{{\cal O}024}}\rightarrow H_{\underline{{\cal O}02}}$,
$H_{\overline{{\cal O}13}}\rightarrow H_{\overline{{\cal O}1}}$, etc., and we have 
{\footnotesize
\[
H_{\underline{{\cal O}02}} \!=\!
\left(\!\!
\begin{array}{ccc}
  \underline{A0}&\underline{B0}\!\!+\!\!\underline{B2} & 0 \\
\underline{B0}^{\dagger} & \underline{A0}&\underline{B0}\!\!+\!\!\underline{B2}\\
0& \underline{B0}^{\dagger} & \underline{A0}
  \end{array}\!\!\right),\qquad
H_{\overline{{\cal O}1}} \!=\!
\left(\!\!
\begin{array}{ccc}
  \overline{A1}\!\!+\!\!\overline{D1}&\overline{B1}& 0 \\
  0 & \overline{A1}\!\!+\!\!\overline{D1} &\overline{B1}\\
0& 0 & \overline{A1}\!\!+\!\!\overline{D1}
  \end{array}\!\!\right),
\]
}
{\footnotesize
\[
H_{\overline{{\cal O}2}} =
\left(\!
\begin{array}{ccc}
0& \overline{B2}_{3Tr}&0\\  
0&0& \overline{B2}_{3Tr}\\
0&0&0
\end{array}\!
\right),\qquad
H_{\widetilde{{\cal O}1}} =
\left(\!
\begin{array}{ccc}
  \widetilde{A1}\!\!+\!\!\widetilde{D1}&\widetilde{B1} & 0 \\
  \widetilde{B1}^{\dagger}& \widetilde{A1}\!\!+\!\!\widetilde{D1}
  &\widetilde{B1}\\
  0& \widetilde{B1}^{\dagger}  & \widetilde{A1}\!\!+\!\!\widetilde{D1}
  \end{array}\!
\right)\!\!,
\]
}
replacing the block matrices of Eq.~(\ref{HamStructure}).
This will simplify future studies of the finite $N$ theory, as the
number of matrix elements relative to asymptotic eigenstates is reduced
substantially. In particular, the only trace-diagonal finite $N$
correction is $\underline{B2}$.


\section{Conclusions}

In this short note we put forth evidence that the insensitivity of
two-dimensional adjoint QCD to the number of colors $N$
is a natural consequence of the
requirement that the theory be finite. Namely, singular terms creeping
in at finite $N$ must cancel, since there are no finite $N$ contribution
to the contractions terms which guarantee a well-behaved, principal value
behavior of the bound state integral equation at large $N$.

The cancellation of 
terms is complete for the trace-number and parton-number
conserving terms. For the trace-number violating terms ($A1$ and $D1$) we
observe some cancellations in the matrix elements relative to a (naive)
single-momentum basis. While it is clear that these singular terms do
have to cancel in the full theory, it is not clear {\em how} this is
realized. Numerical calculations \cite{KlebFiniteN} point to a weak $1/N^2$
behavior of the spectrum. Having ruled out a direct $1/N^2$ behavior
of the trace-diagonal interactions, we can look at the block structure of the 
Hamiltonian, Eq.~(\ref{BlockStructure}), to see that iterating the
trace off-diagonal $1/N$ interactions $A1$ and $D1$ yield such contributions.
Since the singular parts have to vanish, this leaves the regular part of $A1$.
However, also some of the surviving parton-number violating $B$ terms are $1/N$.
It is thus an open question where the weak dependence of the spectrum
on $N$ comes from\footnote{We have narrowed down the choices in this note,
  of course.}. A full, continuous basis-function calculation
\`a la eLCQ
would be a welcome alternative to the discrete calculations in
\cite{KlebFiniteN}, but is beyond the scope of the present note.
On the other hand, we have shown that the {\em physical reason}
the theory is 
insensitive to the number of colors in its gauge group is its singularity
structure, and that additional simplifications are due to the symmetries
of the theory. This restriction of $N_c$ dependence by bound-state physics
or symmetry is likely {\em not} limited to adjoint QCD$_2$, and it would be
interesting to uncover similar restrictions in other theories.  


\section*{Acknowledgments}

I am grateful for support by the Simon's Foundation to
travel to the 
{\em Collaboration on Confinement and QCD Strings} annual meeting;
discussions with conference participants inspired the present
work. In particular, encouragement by Igor R.~Klebanov to pursue the
current line of work is greatly appreciated.
Thanks to the Ohio State University's Physics 
Department for hospitality while this work was being completed.
Additional discussions with Igor R.~Klebanov and Ofer Aharony 
at the late stages of the paper are acknowledged.


\end{document}